\documentclass[preprint,12pt]{aastex}
\begin{document}
\newcommand{\Halpha}{H$\alpha$}
\newcommand{\ie}{{i.e.}}
\newcommand{\eg}{{e.g.}}
\newcommand{\IRAS}{{\it IRAS}}
\newcommand{\DIRBE}{{\it DIRBE}}
\newcommand{\IRDI}{{\it IRAS/DIRBE}}
\newcommand{\COBE}{{\it COBE}}
\newcommand{\CODI}{{\it COBE/DIRBE}}
\newcommand{\ROSAT}{{\it ROSAT}}
\newcommand{\AXAF}{{\it AXAF}}
\newcommand{\EUVE}{{\it EUVE}}
\newcommand{\NHoo}{{N_{\rm H}}}
\newcommand{\NHxo}{{N_{\rm x}}}
\newcommand{\NHIo}{{N_{\rm HI}}}
\newcommand{\NHII}{{N_{\rm HII}}}
\newcommand{\NHgx}{{N_{\rm G}}}
\newcommand{\NHcf}{{N_{\rm cf}}}
\newcommand{\acm}{cm$^{-2}$}
\newcommand{\as}{s$^{-1}$}
\newcommand{\Hmol}{H$_{2}$}
\newcommand{\pp}{\phn}
\newcommand{\ppp}{\phn\phn}
\newcommand{\pppp}{\phn\phn\phn}
\newcommand{\pq}{\,$\pm$\,}
\newcommand{\pd}{\phn\phn\phn\,---}
\newcommand{\plong}{\hspace{10pt}}
\newcommand{\gte}{$\infty$\phn}
\newcommand{\ZY}{0.3,1.0}
\newcommand{\ZZ}{0.5,1.0}
\newcommand{\Msun}{$M_{\odot}$}
\newcommand{\ayr}{y$^{-1}$}
\newcommand{\ARAA}[2]{ARA\&A, #1, #2}
\newcommand{\ApJ}[2]{ApJ, #1, #2}
\newcommand{\ApJL}[2]{ApJL, #1, #2}
\newcommand{\ApJSS}[2]{ApJS, #1, #2}
\newcommand{\AandA}[2]{A\&A, #1, #2}
\newcommand{\AandASS}[2]{A\&AS, #1, #2}
\newcommand{\AJ}[2]{AJ, #1, #2}
\newcommand{\BAAS}[2]{BAAS, #1, #2}
\newcommand{\ASP}[2]{ASP Conf.\ Ser., #1, #2}
\newcommand{\JCP}[2]{J.\ Comp.\ Phys., #1, #2}
\newcommand{\MNRAS}[2]{MNRAS, #1, #2}
\newcommand{\N}[2]{Nature, #1, #2}
\newcommand{\PASJ}[2]{PASJ, #1, #2}
\newcommand{\RPP}[2]{Rep.\ Prog.\ Phys., #1, #2}
\newcommand{\tenup}[1]{\times 10^{#1}}


\title{On the Internal Absorption of Galaxy Clusters}
\author{John S.\ Arabadjis and Joel N.\ Bregman}
\affil{University of Michigan}
\affil{Dept.\ of Astronomy}
\affil{Ann Arbor, MI 48109-1090}
\affil{jsa@astro.lsa.umich.edu}
\affil{jbregman@astro.lsa.umich.edu}

\keywords{galaxies: clusters -- ISM: general -- X-rays: ISM}

\begin{abstract} 

A study of the cores of galaxy clusters with the Einstein SSS indicated
the presence of absorbing material corresponding to $10^{12}$ \Msun\ of cold
cluster gas, possibly resulting from cooling flows.  Since this amount of
cold gas is not confirmed by observations at other wavelengths, we
examined whether this excess absorption is present in the \ROSAT\ PSPC
observations of 20 bright galaxy clusters.  For 3/4 of the clusters,
successful spectral fits were obtained with absorption due only to the
Galaxy, and therefore no extra absorption is needed within the clusters, in
disagreement with the results from the Einstein SSS data for some of the
same clusters.  For 1/4 of the clusters, none of our spectral fits was
acceptable, suggesting a more complicated cluster medium than the
two-temperature and cooling flow models considered here.  However, even for
these clusters, substantial excess absorption is not indicated.

\keywords{galaxies: clusters: general -- ISM: general --- X-rays: ISM}

\end{abstract}

\section{Introduction} 

A central consequence of the cooling flow model for galaxy clusters is that
cool gas is deposited in the central 200 kpc region at a rate that is
typically 30-300 \Msun\ y$^{-1}$ \citep{wjf,af}.  Although this
model is consistent with a wealth of X-ray data, there has been considerable
skepticism about the validity of this picture because of the difficulty in
finding the end state of this cooled gas.  The gas does not form stars with a
normal initial mass function, so either star formation is heavily weighted to
low mass stars, the material does not form stars but remains as cooled gas, or
the cooling flow model is incorrect.  Consequently, there was considerable
excitement when X-ray observations claimed to discover large amounts of cooled
gas in galaxy clusters with approximately the masses expected from a long-lived
cooling flow \citep{wfjma} (hereafter WFJMA).  They used Einstein SSS data for
21 clusters, corrected for a time-dependent ice build-up, and their spectral
fits yielded an absorption column which they compared to the Galactic value
obtained from the large-beam Bell Labs survey \citep{sgwblhh}.  About half of
the clusters (12/21) had X-ray absorption columns in excess of the Galactic HI
column by at least $3\sigma$, and the excess was correlated with the deduced
rate of cooling gas.  The mass of absorbing gas within the cluster was
determined to be $3\tenup{11}-10^{12}$ \Msun, which is approximately the amount
of cooled gas that would be produced by a cooling flow over its lifetime.

The WFJMA study led to searches at other wavelengths for cold gas in cooling
flow clusters, since $10^{11}-10^{12}$ \Msun\ of HI or \Hmol\ would be easily
detected if its properties were similar to Galactic gas.  Observational
searches for HI usually yielded upper limits \citep{jaf87,jaf91,dvo,ogb}, and
when HI was detected, it was typically two orders of magnitude lower than the
expected HI mass \citep{jaf90,mob,ngjh,hjn}.  One concern was that the HI might
have a velocity dispersion similar to the cluster, making it difficult to
detect in narrow bandwidth studies.  However, a recent wide bandwidth search
for HI rules out such emission, typically at a level of $5\tenup{9}$ \Msun\
\citep{opk}.

Searches for molecular hydrogen have often focused on emission or absorption
from CO millimeter lines, which have led to stringent upper limits
\citep{mj,obmts,bd,bwrhl}.  Recently, searches have employed
the \Hmol\ infrared lines, usually the \Hmol\ (1-0)S(1) line, and emission
has been detected in a few cases \citep{jb,frrsw}.  In their analysis of the
detections, \citet{jb} deduce masses that are about $10^{10}$ \Msun, still
inadequate by two orders of magnitude to be in agreement with the X-ray
observations.

Given the limits on HI and \Hmol, theoretical investigations have examined
whether the gas could be hidden in a form that would be difficult to detect.
The work of \citet{dft} and of \citet{ffj} indicated that the gas might
be difficult to detect, with the most likely form being very cold molecular
gas (near 3K).  However, \citet{vd} argue that the material is unlikely to be
this cold and that the X-ray absorbing material would not have evaded detection
if it were in the form of HI or \Hmol. This agrees with the modeling of
\citet{obmts}, and the detection of the infrared \Hmol\ lines shows that some
of the molecular gas must be warm \citep{jb}.  The theoretical models suggest
that it would be difficult to hide cold gas from detection, although perhaps
not impossible.

This apparent conflict between the WFJMA result and data at other wavebands
raises the concern that there might be a problem with the SSS X-ray
observations.  A different group \citep{wdhh} studied four of the same
clusters as WFJMA using SSS data supplemented by GINGA data as part of a study
of abundance gradients in clusters.  \citet{wdhh} found that the amount of
X-ray absorbing material depended upon various assumptions about the spectra,
such as including a cooling flow in the modeling.  Also, increasing the ice
parameter for the SSS data would lead to a decrease in the X-ray absorbing
column.  In most cases, these changes could reduce but not eliminate an X-ray
absorbing column in excess of the Galactic $\NHIo$ value.  A direct conflict
with the WFJMA work was presented by \citet{t}, who used data from several
instruments on the Einstein Observatory and found that toward M87, no
additional X-ray absorption was required beyond the Galactic $\NHIo$ column.

The \ROSAT\ PSPC spectra should provide a strong test of this extra absorption
since it has good sensitivity across the energy band where the absorption
occurs.  For the clusters where the Galactic $\NHIo \lesssim 5\tenup{20}$
\acm\ and that have claimed excess X-ray absorption, such as M87, the Virgo
Cluster, Abell 1795, Abell 2029, Abell 2142, and Abell 2199, no excess
absorption is required by the PSPC data \citep{bhen,henb,lmblhs,swm,ssj}, in
direct conflict with the work of WFJMA.  Also, PSPC spectra of other cooling
flow clusters, such as Abell 401 and Abell 2597, fail to show excess absorption
\citep{henb,sm}.

It is important to note that most of these spectral fits are for a single
temperature within an annulus or region.  Models with cooling flows can
naturally accommodate considerable internal absorbing material because these
models produce soft emission (from the production of cooling gas), which can
be reduced through absorption in order to agree with the observed spectrum
(\eg, \citet{ws}).  A particularly clear illustration of that is given by
\citet{ssj}, who show that no excess absorption is required for either
single-temperature models or cooling flow models without reheating, but that
excess absorption can occur in the center for a cooling flow model with a
partial covering screen.  A somewhat different approach is taken by \citet{af}
who use PSPC color maps along with a deprojection technique to fit cooling
flows plus internal absorption to nearly all of their galaxy clusters.  They
can achieve agreement with WFJMA when they adopt a partial covering model for
the absorption.  The evidence suggests to us that excess absorption can be
accommodated but is not required for successful spectral fits of clusters along
lines of sight where the Galactic $\NHIo \lesssim 5\tenup{20}$ \acm.

The situation is different along sight lines with higher Galactic column
densities, where excess columns are reported even for isothermal fits to the
data.  \citet{is} observed 2A0335+096, which has a Galactic
$\NHIo = 1.7\tenup{21}$ \acm\ and found an excess of $0.6-1.2\tenup{21}$ \acm,
depending upon the type of fit.  A similar result is found by \citet{afjwdes},
who observed Abell 478 and found an excess of $0.7-1.7\tenup{21}$ \acm\
compared to the Galactic $\NHIo = 1.4\tenup{21}$ \acm.  An important aspect of
these studies is that the excess absorption occurs both inside and outside of
the cooling flow core.

Of direct relevance to this discussion is our recent study where we used the
non-central regions of bright clusters to measure absorption columns for
comparison with Galactic $\NHIo$ and $\NHII$ data (\citet{ab_a}, hereafter
AB).  The motivation was that the bright isothermal parts of galaxy clusters
were ideal background light sources with particularly simple spectra, so
absorption columns could be determined to high accuracy.  We found that for
X-ray absorption columns $< 5\tenup{20}$ \acm, the only absorption necessary
was due to Galactic $\NHIo$. However, for the seven clusters with higher
Galactic column densities, excess absorption was detected in every case and we
attribute this excess to \Hmol\ in the Galaxy, a result that is consistent with
Copernicus \Hmol\ studies \citep{sbdb}.  As part of our investigation, we
developed software to incorporate the most recent values of the He absorption
cross section, to which the results are are somewhat sensitive.  Here we extend
the techniques that we developed to study the centers of these 20 bright
clusters with the goal of determining whether excess absorption is required,
and whether it is statistically different than the absorption seen in the
non-central parts of galaxy clusters.

\section{Method and Sample Selection} 

For this investigation we use the cluster sample studied in AB
(Table~\ref{tab:sample}).  These clusters were chosen to fulfill several
criteria:  they must be sufficiently bright such that there were enough
photons in each archived observation to constrain the spectral models; they
must be well-studied so that we minimize the number of free parameters in the
models; they must lie out of the plane of the Galaxy so that opacity
corrections in the corresponding HI columns are minimal.  The data consisted
of \ROSAT\ PSPC observations taken from the archives at the HEASARC.  Standard
packages (\ie, the PCPICOR suite in FTOOLS) were used to correct for spatial
and temporal gain fluctuations in the \ROSAT\ detectors (PSPC B and C; see
\citet{bbp}).  Spectra were usually taken from 3-6$^{\prime}$ and
6-9$^{\prime}$ annuli centered on the emission center of each cluster (but well
outside of any possible cooling flows), over the energy range 0.14-2.4 keV
(avoiding the softest channels where the calibration may be unreliable -- see
\citet{bbp}; \citet{stgy}), and modelled using both XSPEC \citep{arnx} and PROS
\citep{cdmors}.  Background spectra with point sources removed were generally
taken from annuli with widths between 2-4$^{\prime}$ and radii between
$15^{\prime}$ and $20^{\prime}$, and events were binned to ensure a minimum of
20 photons for each channel used in the fitting process.  Each resulting
background-subtracted spectrum was modelled as a single-temperature thermal
plasma (model MEKAL in XSPEC; \citet{mgv}; \citet{mlv}; \citet{ar}; \citet{k})
at a fixed temperature and redshift \citep{wjf} and metallicity (0.3 solar) and
with variable Galactic absorption and spectral normalization.  As mentioned
above, we have replaced the neutral helium cross sections of \citet{bcm} in
XSPEC with the more recent calculations of \citet{ysd}, and set the helium
abundance He/H = 0.10 (see discussion in AB and in \citet{ab_b}).

In the present study our goal is to determine if we can model the emission
from a $3^{\prime}$ disk at the emission center using the same Galactic
absorption column as that derived from X-ray fits to the outer regions,
rendering an absorption component local to the cluster unnecessary.  For each
cluster we try to fit a single-component thermal plasma at the same
temperature, redshift, and metallicity ($T$, $z$, and $Z$) as the models of
AB.  This leaves only one free parameter, the spectral normalization.

Many galaxy clusters appear to exhibit a spatial metallicity gradient,
especially those containing cooling flows
\citep{fp,pbcew,ktt,mkattmh,efmotxy,hmlmfm}.  Roughly speaking, the metallicity
ranges from 0.3-0.5 in the outer regions to approximately solar at the cooling
flow center \citep{es,fofcimmy,mlatfmkh,hmlmfm,af}.  Allowing the metallicity
to vary in our models often produces implausible values, however, with
$Z\sim 4-20$.  This seems to be the result of a competition between metallicity
and absorption to reproduce the sharpness of the spectral peak at 1 keV; \ie,
the feature can be sharpened either by increasing the Galactic column or
increasing the metallicity.  For our models the simplest solution is to use 0.3
for the thermal plasma metallicity, and if the fit obtained is unacceptable
(\eg, one in which the reduced chi-squared $\chi^2_r$ of the fit exceeds 1.26
for 187 degrees of freedom, indicating a probability of less than 1\%), we
increase it to 0.5.  For the cooling flows we adopt $Z=1$.  Our choice of
metallicity does not have a large effect upon our derived absorption columns,
although it should be noted that the effect is somewhat greater for the
resulting mass deposition rates, reducing them by 10-20\% when $Z$ is increased
to 0.5 from 0.3.

If increasing the metallicity to 0.5 fails to improve the fit, we add a
second thermal plasma at the same redshift and metallicity.  This adds two
free parameters, the temperature and normalization of the second emission
component.  If this results in an unacceptable fit, we allow the aborption
column to vary.  The models used for each cluster are shown in
Table~\ref{tab:fits}.

In order to facilitate a comparison with the WFJMA results we also run cooling
flow models (\ie, a thermal plasma plus emission from a cooling flow) for each
cluster.  We use the model of \citet{ms} (\ie\ the CFLOW routine in XSPEC) for
the cooling flow component, as did WFJMA.  The addition of the cooling flow
adds a number of free parameters: the temperature range $T_{lo}$ and $T_{up}$
of the emitting material, the slope $\alpha$ of the power law emissivity
function, and the cooling flow mass deposition rate $\dot{M}$, as well as the
redshift and metallicity.  In these models we set $T_{up}$ to the temperature
of the thermal plasma component (as was done in the WFJMA study), leaving
$T_{lo}$ a free parameter.  We note that $T_{lo}$ could have been set to an
arbitrarily low value (where the gas no longer contributes to emission in the
soft band), but allowing it to vary produced slightly better fits in a few
instances.  In any case, the differences in the fits produced by the two
methods are quite small.  We assume an emission measure that is proportional to
the inverse of the cooling time at the local flow temperature, corresponding to
$\alpha=0$.  The cooling rate $\dot{M}$ is left as a free parameter.

For each cluster we fit several cooling flow models which differ in their
approach to the absorption.  The first model holds the intervening column
constant, at the Galactic value of AB.  The second model allows the column to
vary.  It could be argued that any additional absorption seen in this model is
not truly ``local'', however, since it is manifest only as an increase in the
Galactic column.  Therefore we run a third model wherein the Galactic column
is fixed (at a value determined in AB) and a separate, redshifted absorber
covers only the central cooling flow.  It should be noted, however, that such
an approach does not allow for the expected small-scale structure in the
Galactic interstellar medium ($\sim 7\%$ on these scales; AB), nor is the poor
spectral resolution of \ROSAT\ data capable of distinguishing between absorbers
with differing (low) redshifts.

\section{Results} 

Most of the clusters in our sample do not require an extra absorption
component to be modelled successfully.  Model fits for each cluster are shown
in Table \ref{tab:fits}.  Of the 20 clusters in the sample, 12 can be fit with
a one- or two-component model with the intervening column set to the Galactic
$\NHxo$ value, and thus require no extra absorption component.  Figure~\ref{f1}
shows an acceptable model spectrum, convolved with the PSPC instrument profile,
for Coma (Abell 1656), a cluster in the direction of low Galactic absorption.
The model used here consists of one emission component at a temperature of 8.0
keV, with a Galactic column set to $0.60\tenup{20}$ \acm, a value determined
from fits to the X-ray emission more than $6^{\prime}$ minutes from the
emission center.  (A nearby region was determined by AB to have a column of
$0.78\tenup{20}$ \acm.  Both of these values deviate from the 21 cm column of
\citet{hb} by more than the expected 5-7\% -- see AB for a discussion.)
Figure~\ref{f2} shows the fit for Abell 2657, which lies in a direction of
relatively large Galactic column ($\NHxo = 1.13\tenup{21}$ \acm).  This model
also uses a single emission component ($T = 3.4$ keV), with the column set to
the value derived for an annulus 3-6$^{\prime}$.

For the 8 clusters that cannot be fit adequately using $\NHxo$ from AB,
we allow the Galactic column to vary in order to ascertain whether extra
absorption is required.  {\it In no case do we achieve an acceptable fit}
(\ie, $\chi^2_r < 1.26$) {\it by allowing the column density to deviate from
the value obtained using the outer parts of each cluster}.  In three of these
clusters, however, the fits are only marginally unacceptable.  Abell 85
($\chi^2_r = 1.307$; Figure~\ref{f3}) requires absorption about 6\% higher
than the Galactic $\NHxo$ value at a significance of about 1.5$\sigma$, rather
weak evidence for an absorption component local to the cluster.  The best fit
for Abell 496 ($\chi^2_r = 1.273$; Figure~\ref{f4}) requires a Galactic column
which is about 8\% {\it lower} than the nominal $\NHxo$ value at about the
$2.6\sigma$ level.  The brighter of the two emission peaks in Abell 2256 can
be fit equally well using either the Galactic column from AB or by allowing
$\NHxo$ to vary.  In the latter case the resulting column is {\it lower}, but
by less than 3\% (less than $1\sigma$ significance).

The difference between the $\NHxo$ fit in the center and in the outer parts of
each cluster is expected from normal fluctuations of Galactic $\NHoo$ on these
angular scales, which are typically at the 5-7\% level \citep{cd,sfd,ab_a}.
Alternatively, they may be the result of small calibration errors in the PSPC
response matrices \citep{phs}.  Neither Abell 85 nor Abell 496 shows a
systematic fluctuation in its residuals, which would undermine confidence in
the choice of models used, but the nominal uncertainty in each channel is
perhaps too small, artificially inflating the $\chi^2$ value of the fit.  Such
calibration errors probably dominate the $\chi^2$ of the best-fit model for
Abell 1795.  The fit is unacceptable ($\chi^2_r = 3.85$), but the residual
pattern in Figure~\ref{f5} demonstrates the effect of a probable gain offset
below 0.5 keV \citep{phs} coupled with small statistical errors derived from
the large number of counts ($6\tenup{5}$).  The cooling flow model fit of A1795
is of equally poor quality, but the resulting excess column is closer to the
Galactic value (+19\% for the cooling flow versus +29\% for the two-thermal
component model).

Two-component model fits to the remaining 5 clusters are poor, but if they
are physically significant they show the same behavior as the rest of the
sample.  Allowing each of their Galactic columns to vary does reduce the
$\chi^2$ of the fit, yielding a model with a higher column, although the
significance of this is difficult to ascertain due to the poor significance of
the resulting model.  If we assume that these fits are physically significant,
the columns exceed their Galactic values by $\leq38$\% (48\% for the cooling
flow models), which is typically an order of magnitude smaller than the
excesses found by WFJMA (see Table~\ref{tab:comp}).  For example, clusters
displaying absorption above the Galactic value in both studies (Abell 85,
1795, 2029, and 2199), but otherwise do not seem to be unusual in $\NHoo$,
show an excess that is 40 times greater in WFJMA than in the present work.

The models of WJFMA all contain cooling flows, so for completeness we ran
cooling flow models with variable absorption (both Galactic and proximal to
the cooling flow) for the entire sample.  For those models which contain only
a Galactic absorption component (as a free parameter), in no case was a
substantial excess absorption required to model the emission.  We cannot rule
out the presence of a significant quantity of cool gas at the center of cooling
flows, but we stress that a significant excess absorption is not a
{\it required} feature of these spectra.  Of the internally absorbed cooling
flow clusters common to both WFJMA and this study, only one quarter show
significant excess absorption.  The fact that any show significant absorption
is not surprising, since absorption can be invoked to obscure any amount of
cooling flow emission; that only a quarter actually display this behavior
suggests that excess internal absorption is probably not a ubiquitous feature
of these systems.

It is difficult to compare these results with those of \citet{af} since the
methods differ significantly; however, one point is worth mentioning.  The
``color profile'' approach that they adopted used data from 0.4 keV through
2 keV.  In low Galactic column clusters most of the absorption is manifest from
0.2 to 0.4 keV, where our technique is quite sensitive.  For example, they
compute an excess $\NHoo$ of almost 600\% for A2029, whereas our two-component
model is only 11\% larger than the Galactic value (and lower still for our
externally absorbed cooling flow model; see Table~\ref{tab:comp}).

Figures~\ref{f6} and \ref{f7} show a direct comparison between between our
cooling flow models of 2A0335+096 and A0085, respectively, and those of the
WFJMA study.  The first spectrum shown in each figure with its residuals is
the application of the WFJMA model to the \ROSAT\ data.  Each of the model's
two absorption components (the Galactic column and an absorber in proximity to
the cooling flow), plasma temperature, and cooling flow mass deposition rate
are taken from WFJMA, while the plasma normalization is left as a free
parameter.  The emission and absorption physics used in WJFMA, \citet{rs} and
\citet{mm}, respectively, is also used here.  The second spectrum plotted in
each figure is the single absorption component (\ie\ a variable Galactic
column) cooling flow model of this study.  In both cases our fit is
significantly better than WFJMA (2A0335:  $\chi^2_r = 1.11$ vs.\ 2.05; A0085:
$\chi^2_r = 1.15$ vs.\ 8.70).  In the case of 2A0335, we find an excess column
approximately half that of WFJMA.  For A0085, however, it is more than an order
of magnitude lower.

\section{Summary and Conclusions} 

We have examined the centers of 20 X-ray bright galaxy clusters for evidence
of internal absorption by cool gas.  12 of the 20 clusters can be adequately
fit by a one- or two-component model using the Galactic column density
determined though X-ray absorption to the outer regions of each cluster.  None
of the best-fit models of the 8 remaining clusters becomes an acceptable fit
by allowing the absorption to vary, although three of them are borderline
cases (\ie, their reduced chi-squared values are close to the cut-off of 1.26).
Their columns each deviate from the Galactic absorption to the outer parts of
the clusters by 3-8\%, much less than the large deviations found by WFJMA, and
two of these three have a {\it lower} value.  This is consistent with emission
contrasts due to small-scale structure in the Galactic interstellar medium,
therfore no change in $\NHxo$ beyond those expected are seen.  The remaining
cluster centers are not fit successfuly by either the one-component or
two-component models used here, and although allowing their columns to vary
does reduce their $\chi^2$ values, they never reach acceptable levels.
However, if we assume that these best fits yield valid information about
$\NHoo$, the resulting column density increases are only 11-38\%, more than an
order of magnitude below those seen by WFJMA.  At least 3/4 of this sample
require no absorption beyond that expected from the Galaxy.  Cooling flow
models wherein the sole (Galactic) absorption component is left as a free
parameter show excess absorption at least an order of magnitude lower than
those seen in WJFMA.

We suggest that the discrepancy between our work and that of WFJMA is probably
due to the Einstein SSS calibration.  The WFJMA results depend upon the values
chosen for the SSS ice buildup parameters, and although they used the best
available values, there could be significant uncertainties.  The
time-dependent thickness of the ice buildup varied with position on the solid
state detector, producing an extra absorption component (equivalent to
absorption of between $10^{20}$ and $10^{21}$ \acm) that is significantly
larger than many of the columns being measured.  The standard model for the
behavior of the ice buildup attempts to correct for the extra absorption, and
is valid to a low energy cut-off near the oxygen edge at 0.5 keV \citep{mmwau}.
Unfortunately, low and intermediate Galactic columns ($\NHgx \leq 5\tenup{20}$
\acm) are most readily measured in the 0.14-0.5 keV band (AB), limiting
confidence in these measurements.  Although the data no longer {\it require}
extra $\NHxo$, it may be possible to accomodate extra absorbing material in
certain models \citep{ssj,ws}.

We would like to acknowledge financial support from NASA grant NAG5-3247.
We would also like to thank J.\ Irwin and M.\ Sulkanen for many useful
discussions.



\clearpage
\plotone{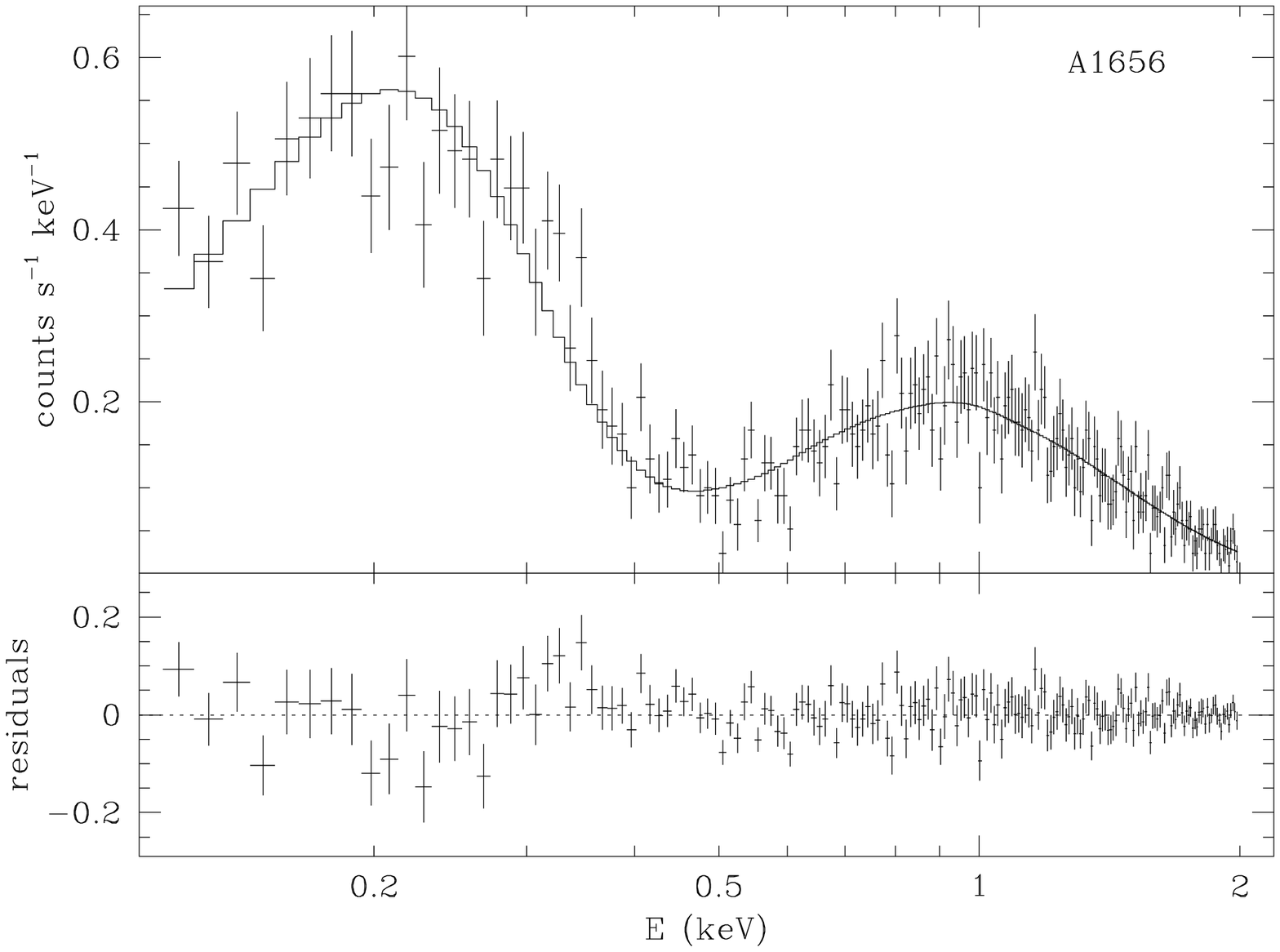}

\figcaption{Model spectrum of Coma (Abell 1656).  The model used is a single
thermal plasma emission component with intervening absorption set to the
Galactic $\NHxo$ value from AB.  The fit is acceptable, with
$\chi^2_r = 1.136$.
\label{f1}}

\clearpage
\plotone{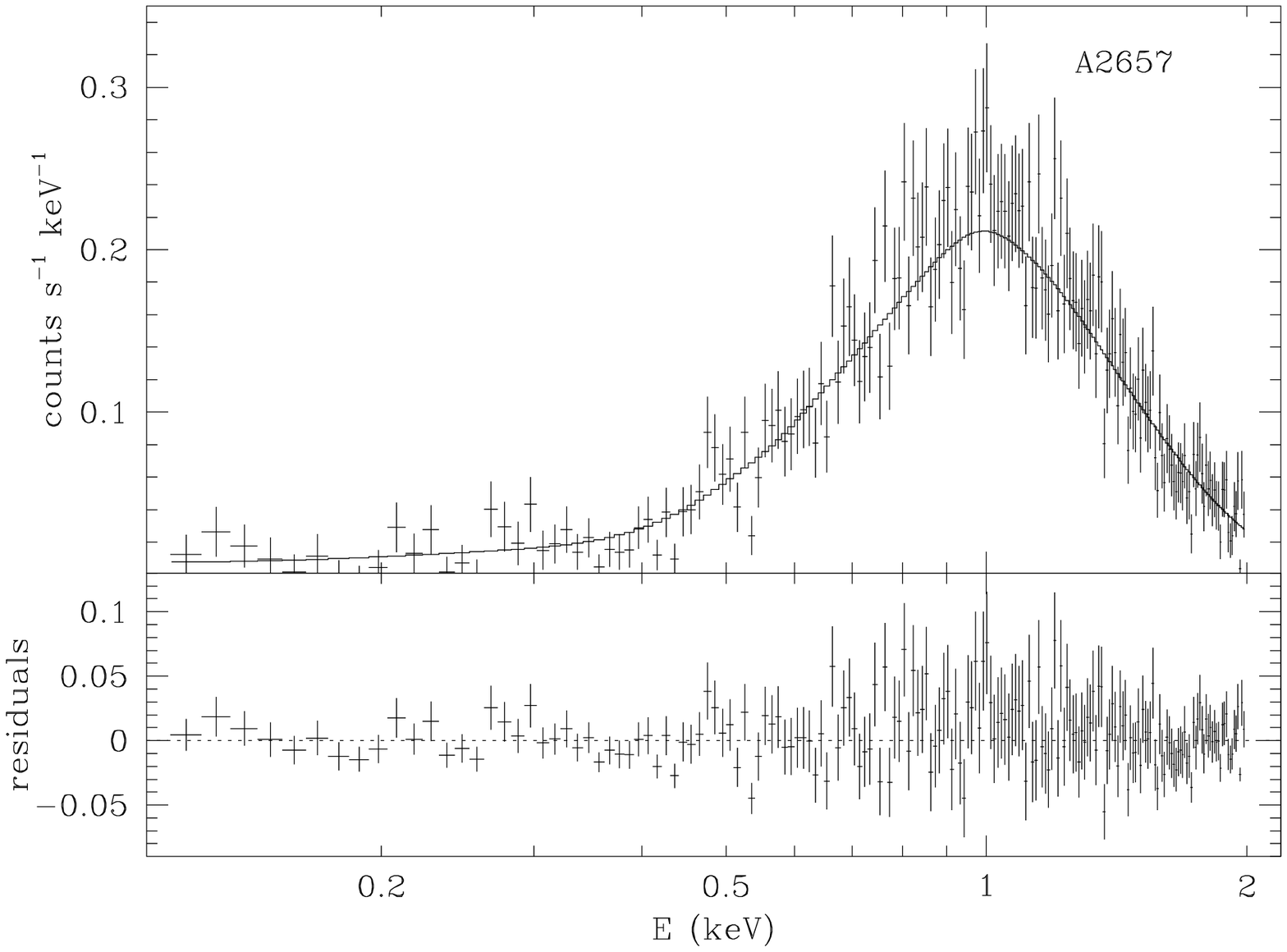}

\figcaption{Model spectrum of Abell 2657.  The model used is a single
thermal plasma emission component with intervening absorption set to the
Galactic $\NHxo$ value from AB.  The fit is acceptable
($\chi^2_r = 1.168$).
\label{f2}}

\clearpage
\plotone{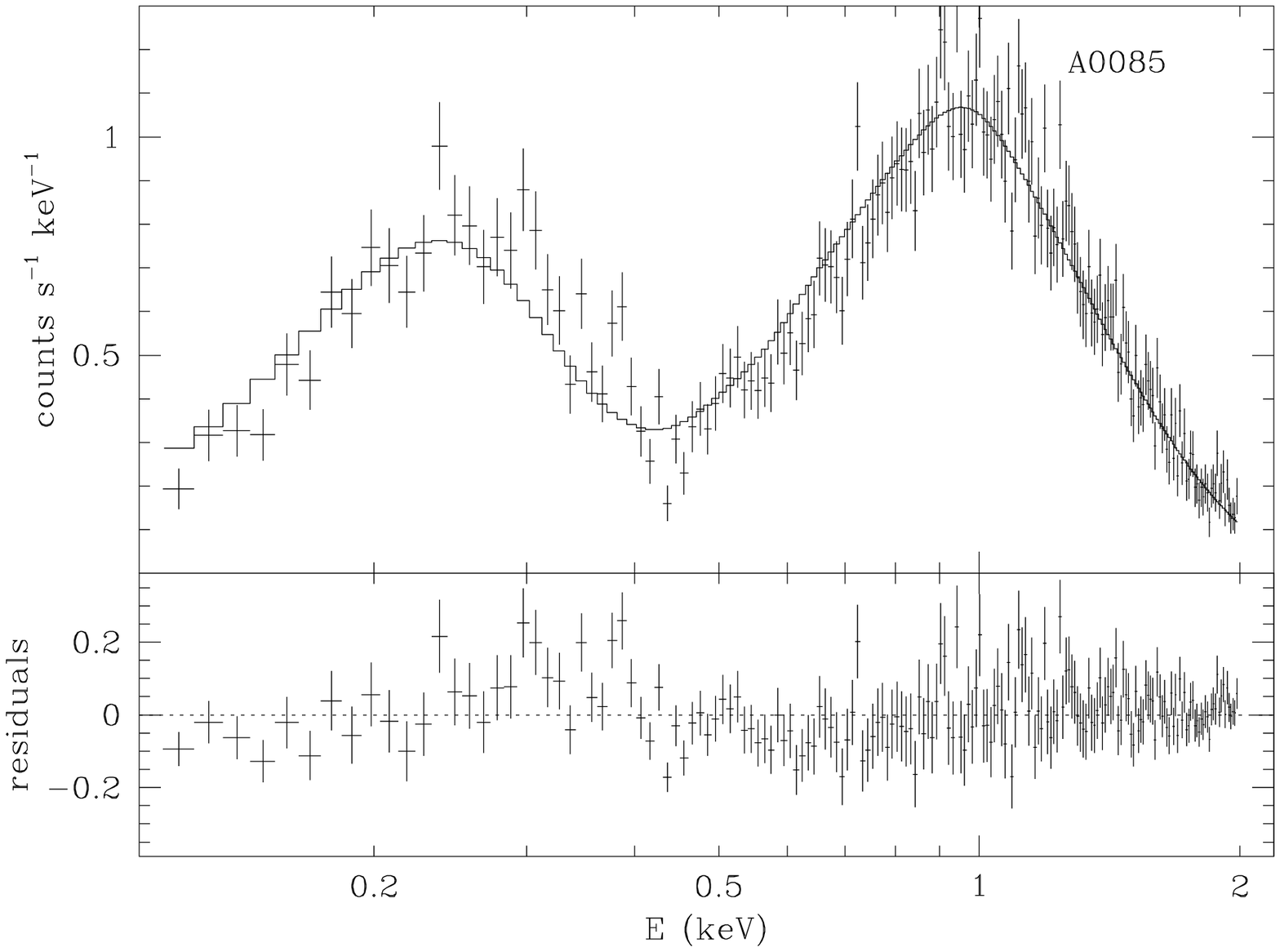}

\figcaption{Model spectrum of Abell 85.  The model used is a two-component
thermal plasma with a variable absorption column.  In this case the column
assumed a value 6\% higher than the Galactic value.  The fit is marginally
unacceptable ($\chi^2_r > 1.26$).
\label{f3}}

\clearpage
\plotone{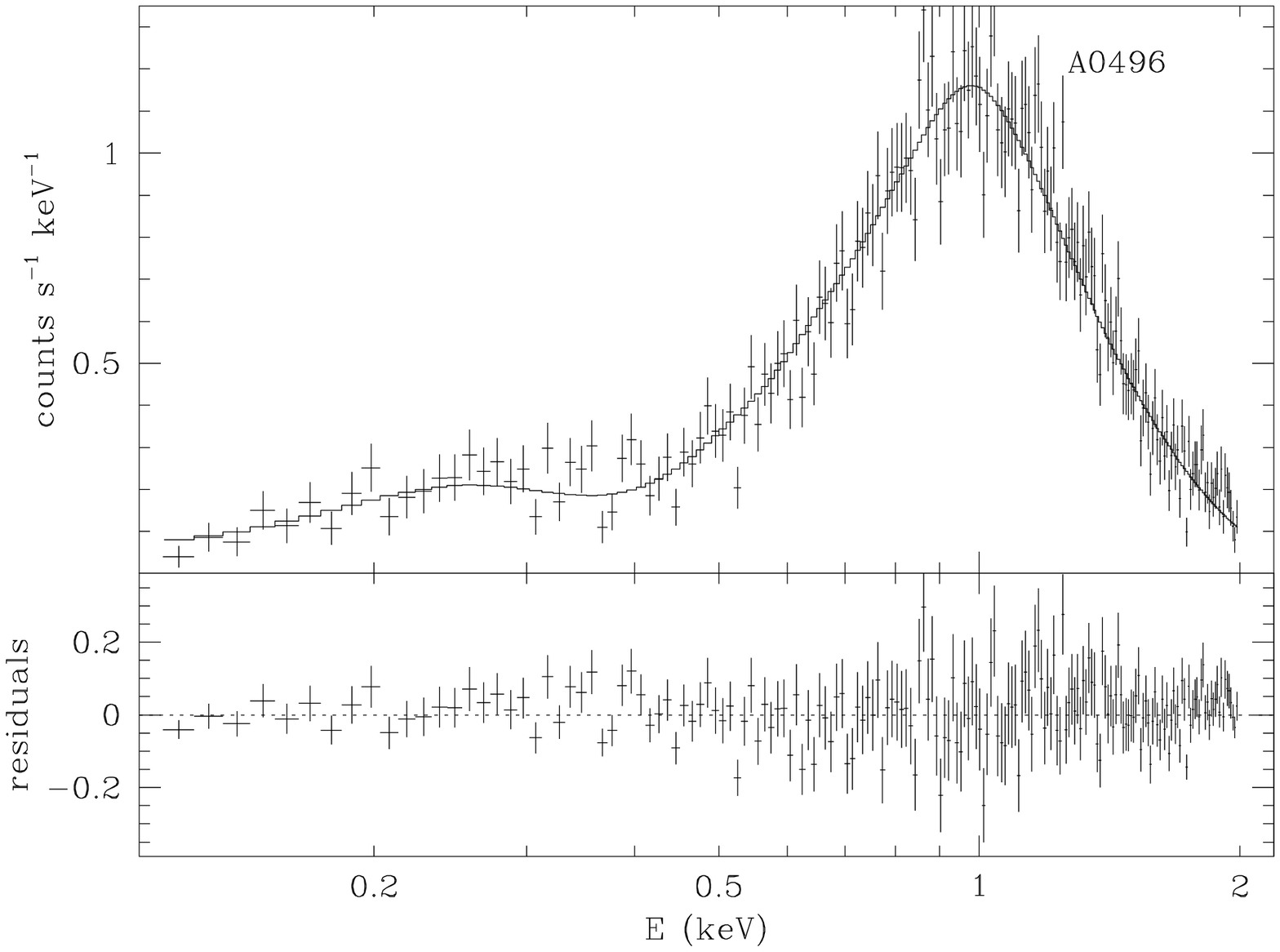}

\figcaption{Model spectrum of Abell 496.  The model used is a two-component
thermal plasma with a variable absorption column.  Here the column assumed
a value 8\% lower than the Galactic value.  The fit is marginally
unacceptable ($\chi^2_r = 1.27$).
\label{f4}}

\clearpage
\plotone{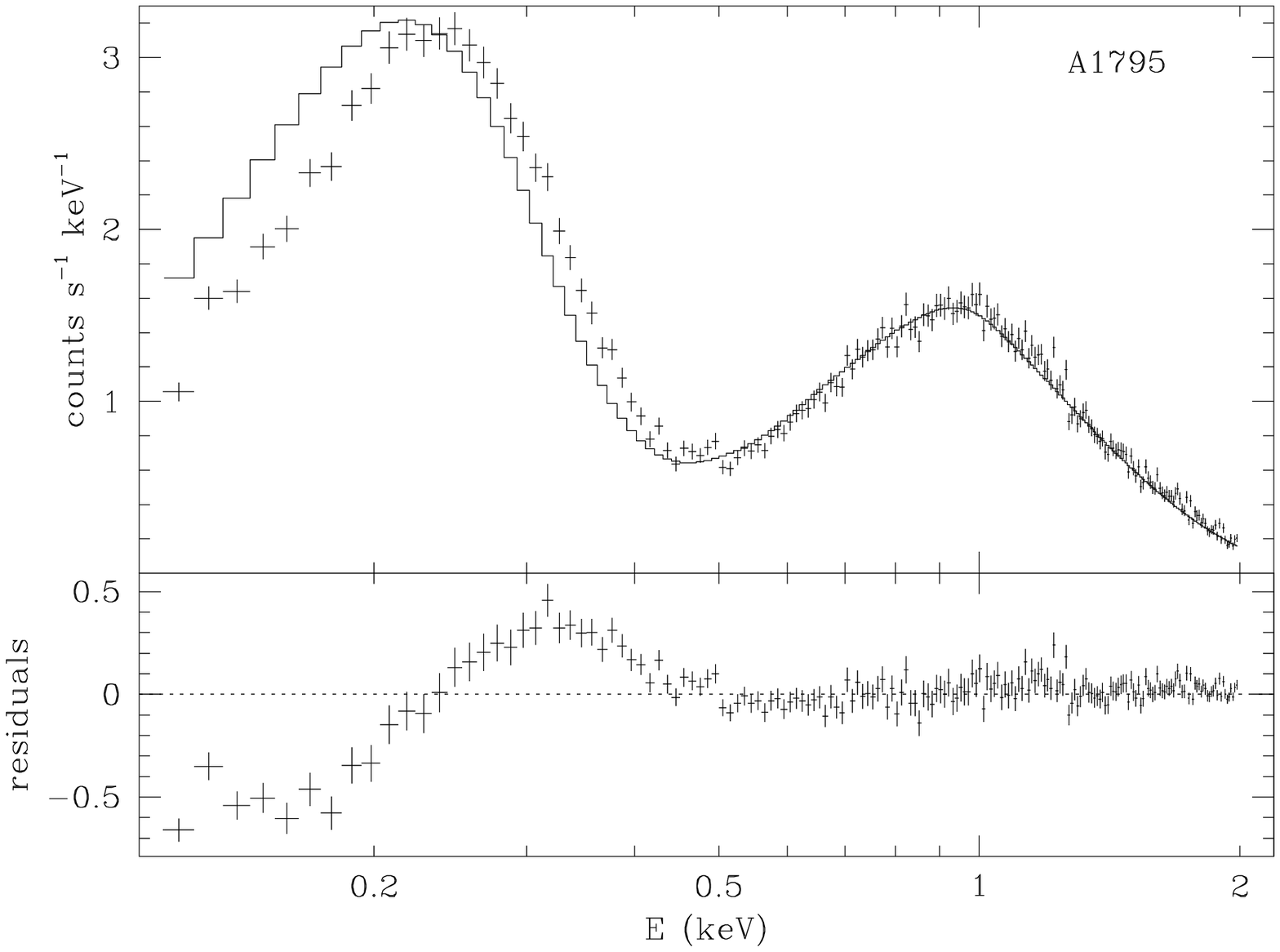}

\figcaption{Failed two-component model spectrum of A1795.  The column here
assumed a value 29\% above the Galactic value.  The systematic errors in the
residuals below 0.5 keV are most likely due to a gain offset.
\label{f5}}

\clearpage
\plotone{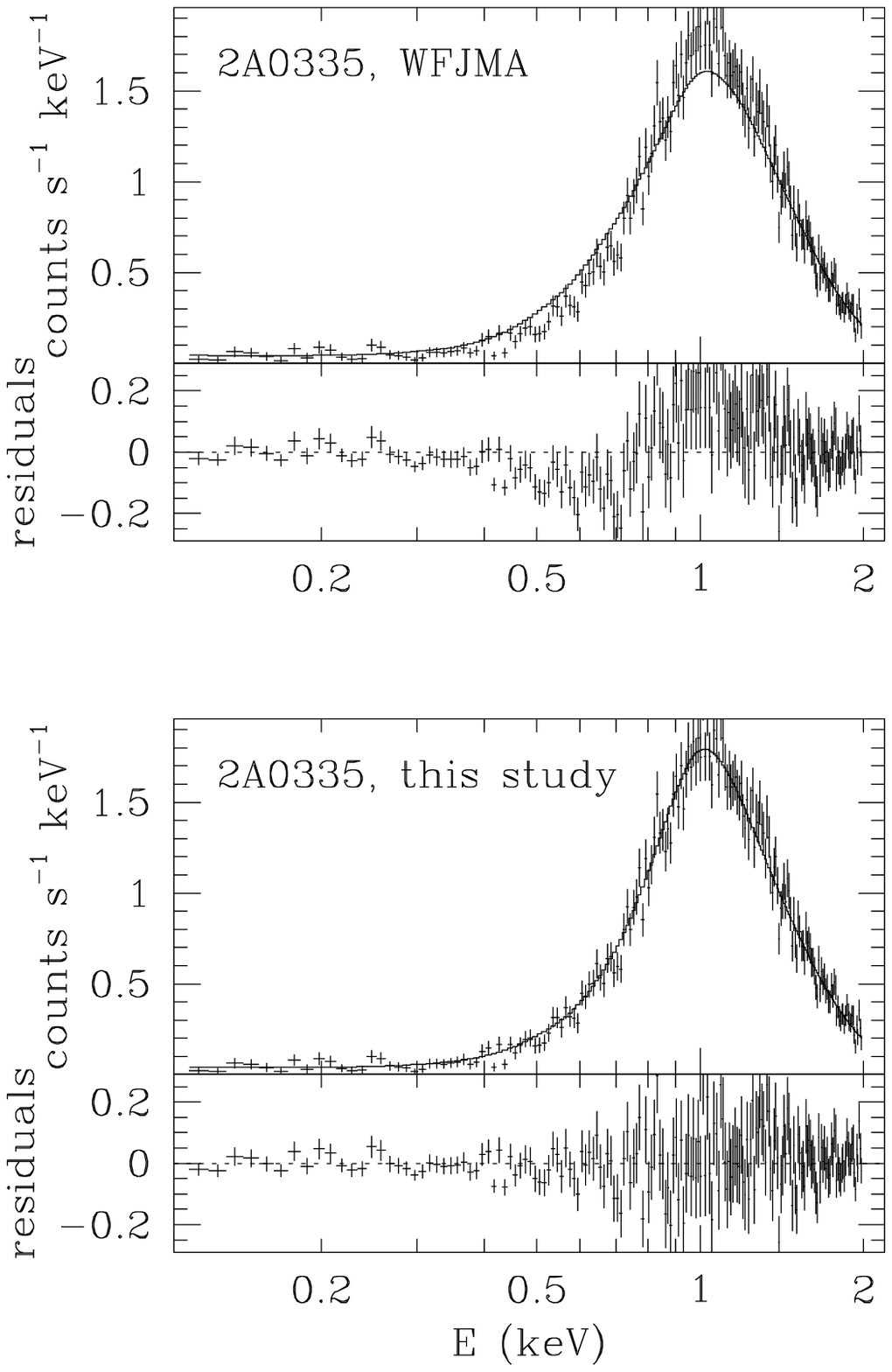}

\figcaption{A comparison of models for the \ROSAT\ spectrum of 2A0335.  The
top spectrum shows the best-fit WJFMA model applied to the data; the bottom
shows the cooling flow and single absorption component model of this study.
\label{f6}}

\clearpage
\plotone{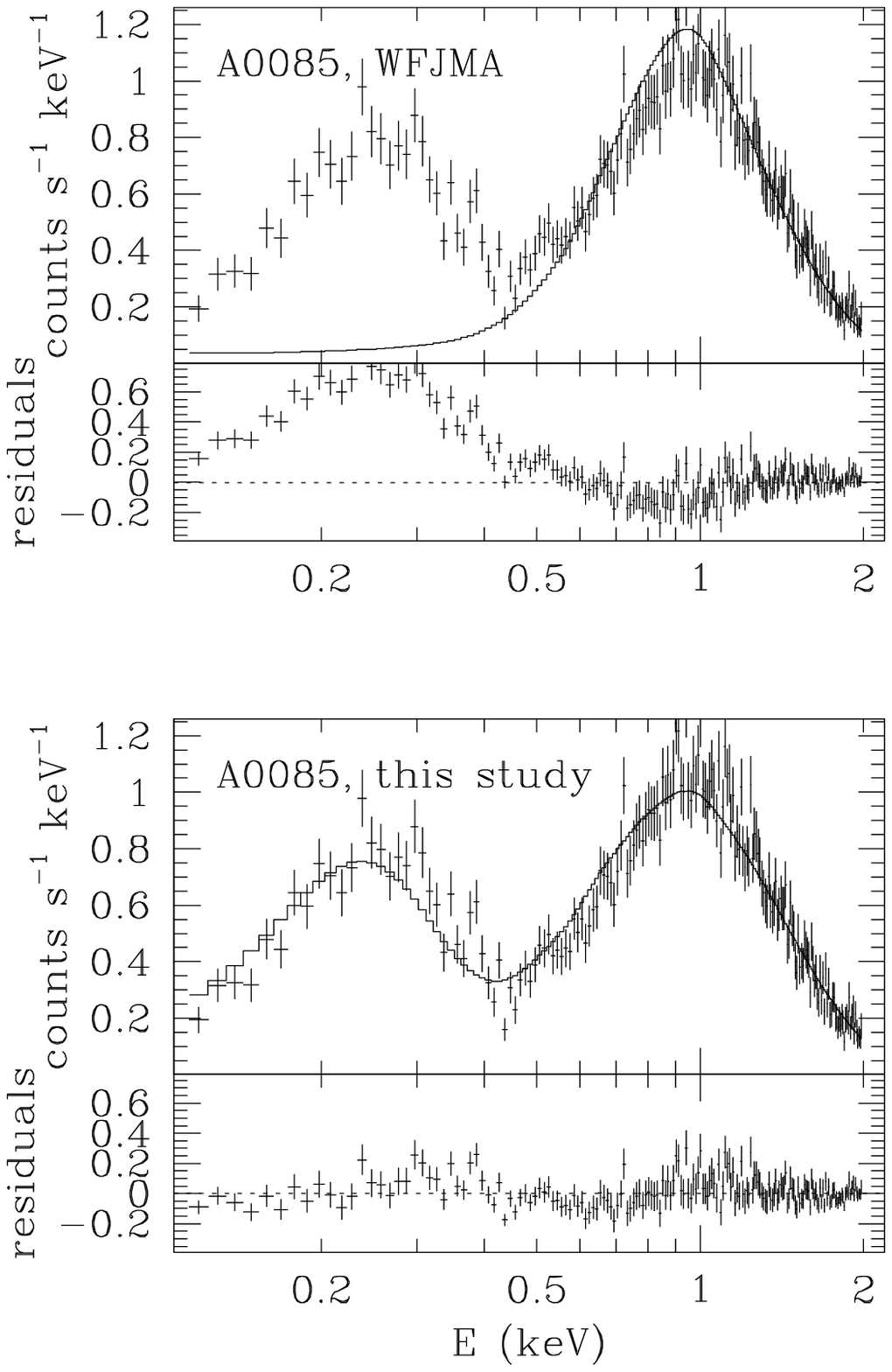}

\figcaption{A comparison of models for the \ROSAT\ spectrum of A0085.  The
top spectrum shows the best-fit WJFMA model applied to the data; the bottom
shows the cooling flow and single absorption component model of this study.
\label{f7}}


\clearpage
\begin{deluxetable}{lrr}
\tablewidth{170pt}
\tablecaption{The 20 galaxy clusters in the sample.
\label{tab:sample}}
\tablehead{
\colhead{cluster}        &
\colhead{{\it l}$^{II}$} &
\colhead{{\it b}$^{II}$}
}
\startdata
 2A0335  & 176.25 & --35.08 \\
  A0085  & 115.05 & --72.08 \\
  A0119  & 125.75 & --64.11 \\
  A0133  & 149.09 & --84.09 \\
  A0401  & 164.18 & --38.87 \\
  A0478  & 182.43 & --28.29 \\
  A0496  & 209.59 & --36.49 \\
  A0665  & 149.73 &  +34.67 \\
  A1060  & 269.63 &  +26.51 \\
  A1651  & 306.83 &  +58.62 \\
  A1656  &  58.16 &  +88.01 \\
  A1795  &  33.81 &  +77.18 \\
  A2029  &   6.49 &  +50.55 \\
  A2052  &   9.42 &  +50.12 \\
  A2142  &  44.23 &  +48.69 \\
  A2147  &  28.83 &  +44.50 \\
  A2163  &   6.75 &  +30.52 \\
  A2199  &  62.93 &  +43.69 \\
  A2256  & 111.10 &  +31.74 \\
  A2657  &  96.65 & --50.30 \\
\enddata
\end{deluxetable}

\clearpage
\begin{deluxetable}{lllcrcccl}
\tablecaption{Model fits to the cluster sample.
\label{tab:fits}}
\tablehead{
\colhead{cluster}                      &
\colhead{$\NHgx$\tablenotemark{[a]}}   &
\colhead{$\NHcf$\tablenotemark{[b,c]}} &
\colhead{Z\tablenotemark{[d]}}         &
\colhead{$T_1$\tablenotemark{[e]}}     &
\colhead{e$_2$\tablenotemark{[f]}}     &
\colhead{$T_2$\tablenotemark{[g,c]}}   &
\colhead{$\dot{M}$\tablenotemark{[h]}} &
\colhead{$\chi^2_r$\tablenotemark{[i]}}
}
\tablenotetext{[a]}{Intervening Galactic hydrogen column density in units of
$10^{20}$ \acm.}
\tablenotetext{[b]}{Column density of separate cooling flow absorption
component in units of $10^{20}$ \acm.}
\tablenotetext{[c]}{\gte indicates that the error in the quantity exceeds the
quantity by a factor of $>$100.}
\tablenotetext{[d]}{Metallicity of emission component(s).}
\tablenotetext{[e]}{Temperature of emission component 1 (keV).}
\tablenotetext{[f]}{Emission component 2:  tp = thermal plasma; cf = cooling
flow; -- = none.}
\tablenotetext{[g]}{Temperature of emission component 2 for the thermal plasma,
or low-temperature cut-off of the cooling flow (keV).}
\tablenotetext{[h]}{Cooling flow mass accretion rate (\Msun/y).}
\tablenotetext{[i]}{Reduced $\chi^2$ value for the model fit.}
%
%
\startdata
2A0335&26.1          & \pd        &0.5& 3.1&--& ---         & ---      &2.292\\
      &26.1          & \pd        &0.5& 3.1&tp& 1.70\pq0.26 & ---      &1.108\\
      &26.1          & \pd        &0.5& 3.1&cf& 0.08\pq\gte & 143\pq15 &1.728\\
      &38.6\pq4.1    & \pd        &0.5& 3.1&cf& 0.35\pq0.01 & 574\pq86 &1.250\\
      &26.1          & 3.3\pq1.9  &0.5& 3.1&cf& 0.90\pq0.37 & 459\pq64 &1.114\\
&&&&&&&&\\
 A0085&\pp2.79       & \pd        &0.5& 6.2&--& ---         & ---      &1.469\\
      &\pp2.79       & \pd        &0.5& 6.2&tp& 1.40\pq0.25 & ---      &1.333\\
      &\pp2.96\pq0.11& \pd        &0.5& 6.2&tp& 1.57\pq0.50 & ---      &1.307\\
      &\pp2.79       & \pd        &0.5& 6.2&cf& 0.08\pq\gte &  20\pq13 &1.462\\
      &\pp2.97\pq0.07& \pd        &0.5& 6.2&cf& 0.36\pq6.27 &  14\pq30 &1.427\\
      &\pp2.79       & 24.2\pq11.6&0.5& 6.2&cf& 0.08\pq\gte &  43\pq14 &1.375\\
&&&&&&&&\\
 A0119&\pp3.10       & \pd        &0.5& 5.1&--& ---         & ---      &0.987\\
      &\pp3.10       & \pd        &0.5& 5.1&cf& 0.08\pq\gte &   0\pq3  &0.994\\
      &\pp3.41\pq0.35& \pd        &0.5& 5.1&cf& 0.35\pq2.80 &   5\pq5  &1.026\\
      &\pp3.10       & 246\pq\gte &0.5& 5.1&cf& 0.09\pq\gte &   0\pq4  &1.000\\
&&&&&&&&\\
 A0133&\pp1.46       & \pd        &0.5& 3.8&--& ---         & ---      &2.056\\
      &\pp1.46       & \pd        &0.5& 3.8&tp& 1.41\pq0.14 & ---      &1.625\\
      &\pp1.73\pq0.09& \pd        &0.5& 3.8&tp& 1.87\pq0.40 & ---      &1.499\\
      &\pp1.46       & \pd        &0.5& 3.8&cf& 0.08\pq\gte &   0\pq10 &2.078\\
      &\pp1.51\pq0.06& \pd        &0.5& 3.8&cf& 1.14\pq0.02 & 164\pq61 &1.514\\
      &\pp1.46       & 94.4\pq\gte 0.5& 3.8&cf& 0.09\pq\gte &   0\pq10 &2.089\\
&&&&&&&&\\
 A0401&12.6          & \pd        &0.3& 7.8&--& ---         & ---      &1.014\\
      &12.6          & \pd        &0.3& 7.8&cf& 0.08\pq\gte &  77\pq23 &1.014\\
      &15.9\pq0.1    & \pd        &0.3& 7.8&cf& 0.08\pq\gte & 102\pq92 &0.973\\
      &12.6          & 7.1\pq10.4 &0.3& 7.8&cf& 0.08\pq\gte & 108\pq107&0.982\\
&&&&&&&&\\
 A0478&37.4          & \pd        &0.5& 6.8&--& ---         & ---      &1.528\\
      &37.4          & \pd        &0.5& 6.8&tp& 1.99\pq0.93 & ---      &1.153\\
      &37.4          & \pd        &0.5& 6.8&cf& 0.35\pq0.01 & 620\pq106&1.223\\
      &40.5\pq2.3    & \pd        &0.5& 6.8&cf& 0.35\pq0.97 & 882\pq178&1.155\\
      &37.4          & 12.0\pq4.1 &0.5& 6.8&cf& 0.08\pq\gte &1197\pq421&1.134\\
&&&&&&&&\\
 A0496&\pp7.02       & \pd        &0.5& 4.7&--& ---         & ---      &1.729\\
      &\pp7.02       & \pd        &0.5& 4.7&tp& 1.83\pq0.30 & ---      &1.300\\
      &\pp6.43\pq0.23& \pd        &0.5& 4.7&tp& 1.58\pq0.30 & ---      &1.273\\
      &\pp7.02       & \pd        &0.5& 4.7&cf& 0.17\pq0.02 &  34\pq7  &1.561\\
      &\pp6.57\pq0.22& \pd        &0.5& 4.7&cf& 0.17\pq1.27 &  34\pq8  &1.538\\
      &\pp7.02       & 0.0\pq0.0  &0.5& 4.7&cf& 0.35\pq0.01 &  44\pq11 &1.534\\
&&&&&&&&\\
 A0665&\pp4.73       & \pd        &0.5& 8.3&--& ---         & ---      &1.025\\
      &\pp4.73       & \pd        &0.5& 8.3&cf& 0.08\pq\gte &  99\pq47 &1.018\\
      &\pp4.88\pq0.16& \pd        &0.5& 8.3&cf& 0.22\pq4.5  & 100\pq96 &1.018\\
      &\pp4.73       & 1.37\pq23.8&0.5& 8.3&cf& 0.08\pq6.8  &  97\pq168&1.017\\
&&&&&&&&\\
 A1060&\pp6.53       & \pd        &0.3& 3.3&--& ---         & ---      &0.864\\
      &\pp6.53       & \pd        &0.3& 3.3&cf& 0.08\pq\gte &   0\pq1  &0.859\\
      &\pp6.24\pq0.35& \pd        &0.3& 3.3&cf& 0.23\pq3.27 &   1\pq1  &0.853\\
      &\pp6.53       & 0.0\pq0.0  &0.3& 3.3&cf& 0.10\pq\gte &   0\pq0  &0.865\\
&&&&&&&&\\
 A1651&\pp1.59       & \pd        &0.5& 7.0&--& ---         & ---      &1.444\\
      &\pp1.59       & \pd        &0.5& 7.0&tp& 1.73\pq0.47 & ---      &1.411\\
      &\pp1.78\pq0.15& \pd        &0.5& 7.0&tp& 3.15\pq3.62 & ---      &1.388\\
      &\pp1.59       & \pd        &0.5& 7.0&cf& 0.28\pq3.31 &  45\pq49 &1.441\\
      &\pp1.59\pq0.74& \pd        &0.5& 7.0&cf& 0.28\pq3.38 &  45\pq49 &1.448\\
      &\pp1.59       &0.02\pq\gte &0.5& 7.0&cf& 0.22\pq\gte &   0\pq25 &1.467\\
&&&&&&&&\\
 A1656&\pp0.597      & \pd        &0.3& 8.0&--& ---         & ---      &1.136\\
      &\pp0.597      & \pd        &0.3& 8.0&cf& 1.02\pq\gte &   0\pq1  &1.148\\
      &\pp0.587\pq0.071& \pd      &0.3& 8.0&cf& 0.41\pq\gte &   0\pq1  &1.154\\
      &\pp0.597      & 0.0\pq0.0  &0.3& 8.0&cf& 1.13\pq\gte &   0\pq2  &1.156\\
&&&&&&&&\\
 A1795&\pp0.909      & \pd        &0.5& 5.1&--& ---         & ---      &4.886\\
      &\pp0.909      & \pd        &0.5& 5.1&tp& 1.16\pq0.10 & ---      &5.106\\
      &\pp1.17\pq0.03& \pd        &0.5& 5.1&tp& 2.98\pq0.94 & ---      &3.848\\
      &\pp0.909      & \pd        &0.5& 5.1&cf& 0.56\pq2.04 &  83\pq42 &4.501\\
      &\pp1.08\pq0.02& \pd        &0.5& 5.1&cf& 1.14\pq2.64 &  32\pq31 &3.848\\
      &\pp0.909      &58.8\pq\gte &0.5& 5.1&cf& 0.11\pq\gte &   0\pq21 &5.807\\
&&&&&&&&\\
 A2029&\pp3.23       & \pd        &0.5& 7.8&--& ---         & ---      &2.009\\
      &\pp3.23       & \pd        &0.5& 7.8&tp& 1.29\pq0.15 & ---      &1.592\\
      &\pp3.59\pq0.09& \pd        &0.5& 7.8&tp& 1.29\pq0.15 & ---      &1.503\\
      &\pp3.23       & \pd        &0.5& 7.8&cf& 0.08\pq6.97 & 132\pq31 &1.911\\
      &\pp3.43\pq0.08& \pd        &0.5& 7.8&cf& 0.14\pq2.88 & 136\pq33 &1.808\\
      &\pp3.23       & 4.02\pq\gte&0.5& 7.8&cf& 0.54\pq\gte &   0\pq74 &2.041\\
&&&&&&&&\\
 A2052&\pp3.10       & \pd        &0.5& 3.4&--& ---         & ---      &1.613\\
      &\pp3.10       & \pd        &0.5& 3.4&tp& 1.72\pq0.23 & ---      &1.215\\
      &\pp3.10       & \pd        &0.5& 3.4&cf& 0.55\pq5.26 &  23\pq33 &1.531\\
      &\pp3.08\pq0.11& \pd        &0.5& 3.4&cf& 0.71\pq2.68 &  24\pq15 &1.525\\
      &\pp3.10       & 0.0\pq0.0  &0.5& 3.4&cf& 0.90\pq0.01 &  70\pq31 &1.424\\
&&&&&&&&\\
 A2142&\pp4.17       & \pd        &0.5&11.0&tp& ---         & ---      &1.283\\
      &\pp4.17       & \pd        &0.5&11.0&tp& 0.08\pq0.13 & ---      &1.179\\
      &\pp4.17       & \pd        &0.5&11.0&cf& 0.08\pq7.41 & 188\pq43 &1.190\\
      &\pp4.45\pq0.17& \pd        &0.5&11.0&cf& 0.15\pq3.46 & 192\pq46 &1.149\\
      &\pp4.17       & 1.17\pq5.87&0.5&11.0&cf& 0.08\pq\gte & 210\pq110&1.152\\
&&&&&&&&\\
 A2147&\pp2.56       & \pd        &0.3& 4.4&--& ---         & ---      &0.497\\
      &\pp2.56       & \pd        &0.3& 4.4&cf& 0.15\pq4.68 &  27\pq12 &0.467\\
      &\pp2.83\pq1.06& \pd        &0.3& 4.4&cf& 0.28\pq2.43 &  28\pq23 &0.473\\
      &\pp2.56       & 0.00\pq0.03&0.3& 4.4&cf& 0.22\pq0.03 &  24\pq15 &0.638\\
&&&&&&&&\\
 A2163&26.4          & \pd        &0.3&13.9&--& ---         & ---      &1.312\\
      &26.4          & \pd        &0.3&13.9&tp& 2.17\pq1.60 & ---      &1.124\\
      &26.4          & \pd        &0.3&13.9&cf& 0.08\pq\gte &1570\pq278&1.137\\
      &26.7\pq0.07   & \pd        &0.3&13.9&cf& 0.35\pq3.03 &1631\pq874&1.139\\
      &26.4         & 1.65\pq10.5&0.3&13.9&cf& 0.08\pq\gte &1759\pq1297&1.139\\
&&&&&&&&\\
 A2199&\pp0.877      & \pd        &0.5& 4.7&--& ---         & ---      &6.661\\
      &\pp0.877      & \pd        &0.5& 4.7&tp& 1.27\pq0.05 & ---      &3.992\\
      &\pp1.21\pq0.03& \pd        &0.5& 4.7&tp& 1.96\pq0.16 & ---      &2.770\\
      &\pp0.877      & \pd        &0.5& 4.7&cf& 0.09\pq0.84 &  24\pq2  &5.300\\
      &\pp1.03\pq0.02& \pd        &0.5& 4.7&cf& 0.35\pq0.01 &  26\pq3  &4.280\\
      &\pp0.877      & 2.34\pq1.27&0.5& 4.7&cf& 0.17\pq1.02 &  14\pq2  &4.705\\
&&&&&&&&\\
 A2256&\pp4.65       & \pd        &0.3& 7.5&--& ---         & ---      &1.367\\
      &\pp4.65       & \pd        &0.3& 7.5&tp& 0.86\pq0.18 & ---      &1.294\\
      &\pp4.52\pq0.19& \pd        &0.3& 7.5&tp& 0.87\pq0.19 & ---      &1.298\\
      &\pp4.65       & \pd        &0.3& 7.5&cf& 0.08\pq9.44 &  14\pq6  &1.307\\
      &\pp4.46\pq2.61& \pd        &0.3& 7.5&cf& 0.08\pq\gte &  21\pq11 &1.293\\
      &\pp4.65       & 0.0\pq0.0  &0.3& 7.5&cf& 0.08\pq0.22 &  15\pq6  &1.311\\
&&&&&&&&\\
 A2657&11.3          & \pd        &0.5& 3.4&--& ---         & ---      &1.168\\
      &11.3          & \pd        &0.5& 3.4&cf& 0.90\pq1.12 &  17\pq8  &1.138\\
      &12.6\pq1.5    & \pd        &0.5& 3.4&cf& 0.71\pq2.07 &  29\pq15 &1.077\\
      &11.3          & 13.4\pq11.1&0.5& 3.4&cf& 0.08\pq\gte &  35\pq31 &1.073\\
\enddata
\end{deluxetable}

\clearpage
\begin{center}
\begin{deluxetable}{lrcrrrrr}
\tablecaption{Excess absorption, internal absorption, and cooling rates in
cluster models.
\label{tab:comp}}
\tablehead{
\colhead{}                            &
\colhead{\small$\Delta\NHxo/\NHgx$}         &
\colhead{\small$\Delta\NHxo/\NHgx$}         &
\colhead{\small$\Delta\NHxo/\NHgx$}         &
\colhead{\small$\NHcf/\NHgx$}               &
\colhead{\small$\dot{M}$}                   &
\colhead{\small$\dot{M}$}                   &
\colhead{\small$\dot{M}$}                   \\
\colhead{}                            &
\colhead{\small WFJMA}                       &
\colhead{\small tp + tp}                     &
\colhead{\small tp + cf}                     &
\colhead{\small tp + acf}                    &
\colhead{\small WFJMA}                       &
\colhead{\small tp + cf}                     &
\colhead{\small tp + acf}                    \\
\colhead{\small cluster}                     &
\colhead{\small\% \tablenotemark{a}}      &
\colhead{\small\% \tablenotemark{b}}      &
\colhead{\small\% \tablenotemark{c}}      &
\colhead{\small\% \tablenotemark{d}}      &
\colhead{\small\Msun/y \tablenotemark{e}} &
\colhead{\small\Msun/y \tablenotemark{f}} &
\colhead{\small\Msun/y \tablenotemark{g}}
}
\tablenotetext{a}{ Change in the X-ray absorption column when it is allowed
to deviate from $N_{\rm 21cm}$ taken from \citet{sgwblhh}.}
\tablenotetext{b}{ Change in the X-ray absorption column when it is allowed
to deviate from $\NHxo$ taken from AB.  The emission is modelled as a one- or
two-component thermal plasma.  A zero indicates that an acceptable fit was
obtained by setting the absorption to the Galactic column from AB.}
\tablenotetext{c}{ Change in the Galactic X-ray absorption column when it is
allowed to deviate from $\NHxo$ taken from AB.  The emission is modelled as a
thermal plasma plus a cooling flow, with the absorption column a free
parameter.}
\tablenotetext{d}{ Absorption column of a separate absorption component
(relative to the Galactic column) which covers only the cooling flow.}
\tablenotetext{e}{ Mass deposition rate in the WFJMA cooling flow.}
\tablenotetext{f}{ Mass deposition rate in the cooling flow models of this
study with variable Galactic absorption.}
\tablenotetext{g}{ Mass deposition rate in the cooling flow models of this
study with variable cluster cooling flow absorption.}
\startdata
\small
2A0335 & $+90\ \pm ^{\phn30}_{\phn30}$  & \ppp0       & $+48 \pm 16$    &
         $ 13 \pm 7\pppp$               & $105\ \pm ^{\phn88}_{\phn66}$ &
         $574 \pm \phn86$               & $459 \pm \phn64$              \\
\small
 A0085 & $+330\ \pm ^{170}_{130}$       & $+6 \pm 4$  & $+6 \pm \phn3$  &
         $ 867 \pm 414\pp$              & $290\ \pm ^{138}_{130}$       &
         $14 \pm \phn30$                & $43 \pm \phn14$               \\
\small
 A0401 & $+190\ \pm ^{110}_{\phn70}$    & \ppp0       & $+26 \pm \phn1$ &
         $ 56 \pm 83\ppp$               & $111\ \pm ^{236}_{111}$       &
         $102 \pm \phn92$               & $108 \pm 107$                 \\
\small
 A0478 & $+300\ \pm ^{\phn60}_{\phn50}$ & \ppp0       & $+8 \pm \phn6$  &
         $ 32 \pm 11\ppp$               & $495\ \pm ^{580}_{424}$       &
         $882 \pm 178$                  & $1197 \pm 421$                \\
\small
 A0496 & $+470\ \pm ^{\phn90}_{\phn60}$ & $-8 \pm 3$  & $-6 \pm \phn3$  &
         $\ppp0 \pm 0\pppp$             & $65\ \pm ^{\phn29}_{\phn23}$  &
         $34 \pm \ppp8$                 & $44 \pm \phn11$               \\
\small
 A1656 & $-100\ \pm ^{670}_{\ppp0}$     & \ppp0       & $-2 \pm 12$     &
         $\ppp0 \pm 0\pppp$             & $16\ \pm ^{\phn29}_{\phn15}$  &
         $0 \pm \ppp1$                  & $0 \pm \ppp2$                 \\
\small
 A1795 & $+730\ \pm ^{270}_{270}$       & $+29 \pm 3$ & $+19 \pm \phn2$ &
         $ 647 \pm 2700$                & $225\ \pm ^{144}_{112}$       &
         $32 \pm \phn31$                & $0 \pm \phn21$                \\
\small
 A2029 & $+580\ \pm ^{160}_{160}$       & $+11 \pm 3$ & $+6 \pm \phn2$  &
         $ 124 \pm 6900$                & $513\ \pm ^{304}_{247}$       &
         $136 \pm \phn33$               & $0 \pm \phn74$                \\
\small
 A2142 & $+340\ \pm ^{\phn80}_{110}$    & \ppp0       & $+7 \pm \phn4$  &
         $ 28 \pm 141\pp$               & $143\ \pm ^{141}_{130}$       &
         $192 \pm \phn46$               & $210 \pm 110$                 \\
\small
 A2147 & $+330\ \pm ^{210}_{180}$       & \ppp0       & $+10 \pm 41$    &
         $\ppp0 \pm 114\pp$             & $28\ \pm ^{\phn25}_{\phn12}$  &
         $28 \pm \phn23$                & $24 \pm \phn15$               \\
\small
 A2199 & $+1560\ \pm ^{220}_{200}$      & $+38 \pm 3$ & $+17 \pm \phn2$ &
         $267 \pm 145\pp$               & $60\ \pm ^{\phn20}_{\phn17}$  &
         $26 \pm \ppp3$                 & $14 \pm \ppp2$                \\
\small
 A2256 & $-100\ \pm ^{240}_{\phn20}$    & $-3 \pm 4$  & $-4 \pm 56$     &
         $\ppp0 \pm 0\pppp$             & $200\ \pm ^{140}_{\phn90}$    &
         $21 \pm \phn11$                & $16 \pm \ppp6$                \\
\enddata
\end{deluxetable}
\end{center}

\end{document}